\documentclass[11pt]{article}
\usepackage{graphicx,rcdefs,subfigure}

\newcommand{\BABARPubYear}    {04}

\newcommand{\BABARConfNumber} {09}
\newcommand{\SLACPubNumber} {10613}

\def\Mmiss{\ifmath{m_\mathrm{miss}}}
\let\mMiss=\Mmiss
\def\Ecms{\ensuremath{E_{\gamma}^\mathrm{CMS}}}
\def\pcms{\ensuremath{p^\mathrm{CMS}}}

\def\bdstardsstar {\ensuremath{\Bo\rightarrow D_s^{*+}D^{*-}}}

\def\Dsphipi {\ensuremath{D_s^+ \rightarrow \phi\pi^+}}
\let\dsphipi=\Dsphipi

\def\dstar {\ensuremath{D^{\ast}}}
\def\dsstar{\ensuremath{D_s^{\ast}}}
\def\dstarm {\ensuremath{D^{\ast-}}}

\def\dsstarp{\ensuremath{D_s^{\ast+}}}

\def\nBB {{\ensuremath{N_{\BBo}}}}

\def\Dsp{\ensuremath{D_s^+}}

\def\bdstardsstar {\ensuremath{\Bz\to\Dss\Dstarm}}
\def\dsphipi{\ensuremath{\Ds\rightarrow\phi\pip}}
\let\Dsphipi=\dsphipi
\def\Kappa{\ifmath{\mathcal{K}}}

\def\nDs{\ifmath{\mathcal{N}_{\Dsnosign}}}

\def\nDsPhipi{\ifmath{\mathcal{N}_{\Dsnosign\to\phi\pi}}}

\def\epem{\ifmath{e^{\sscr +}e^{\sscr-}}}

\def\ustat{\ensuremath{_\mathrm{stat}}}
\def\usyst{\ensuremath{_\mathrm{syst}}}
\def\mmax{\ensuremath{m_\mathrm{max}}}

\input babarsym

\long\def\inst#1{\par\nobreak\kern 4pt\nobreak
    {\it #1}\par\vskip 10pt plus 3pt minus 3pt}

\begin{document}
{\pagestyle{empty}

\begin{flushright}
  \babar-CONF-\BABARPubYear/\BABARConfNumber \\
  SLAC-PUB-\SLACPubNumber \\
  August 2004 \\
\end{flushright}

\par\vskip 4cm
\begin{center}
\Large \bf 
Partial Reconstruction of
\boldmath $B^0\to D_s^{\ast+} D^{\ast-}$ Decays and
Measurement of the $D_s^+\to\phi\pi^+$ Branching Fraction
\end{center}
\bigskip
\begin{center}
\large The \babar\ Collaboration\\
\mbox{ }\\
\today
\end{center}
\bigskip \bigskip
\begin{center}
\large \bf Abstract
\end{center}
We present preliminary results on the branching fractions
$\BrFr(\bdstardsstar)$ and $\BrFr(\Dsphipi)$, based on a data sample of
approximately $124\times10^6$ \BB\ events collected by the \babar\ detector
at the PEP-II $e^+ e^-$ \B--factory.
$\BrFr(\bdstardsstar)$ is measured selecting neutral
\B\ meson decays to the final state $\dstarm\dsstarp$ with partial
reconstruction of the \Dss, in which only the \dstarm\ and the soft photons
from the decay $\Dss \to \Dsp \gamma$ are reconstructed.
The branching fraction product $\BrFr(\bdstardsstar)\cdot \BrFr(\Dsphipi)$
is measured via a complete reconstruction of the whole decay chain.
Comparing these two measurements provides a model-independent
determination of the $\BrFr(\Dsphipi)$ branching fraction.
We obtain the following preliminary results:
\begin{center}
$\BrFr(\bdstardsstar) = (1.85 \pm 0.09 \pm 0.16)\%$ \\
$ \BrFr(\dsphipi)     = (4.71 \pm 0.47 \pm 0.35)\%$
\end{center}
where the first error in each measurement is statistical, the second
systematic.
\vfill
\begin{center}

Submitted to the 32$^{\rm nd}$ International Conference on High-Energy Physics, ICHEP 04,\\
16 August---22 August 2004, Beijing, China
\end{center}

\vspace{1.0cm}

\begin{center}
{\em Stanford Linear Accelerator Center, Stanford University, 
Stanford, CA 94309} \\ \vspace{0.1cm}\hrule\vspace{0.1cm}
Work supported in part by Department of Energy contract DE-AC03-76SF00515.
\end{center}

\newpage
} 

\begin{center}
\small

The \babar\ Collaboration,
\bigskip

%
B.~Aubert,
R.~Barate,
D.~Boutigny,
F.~Couderc,
J.-M.~Gaillard,
A.~Hicheur,
Y.~Karyotakis,
J.~P.~Lees,
V.~Tisserand,
A.~Zghiche
\inst{Laboratoire de Physique des Particules, F-74941 Annecy-le-Vieux, France }
% ...

\end{center}\newpage

\section{INTRODUCTION}
\label{sec:Introduction}
A precise measurement of the branching fraction for
the $\Dsphipi$ mode is important because nearly all \Ds branching 
fractions are determined by normalizing the measurements to
$\BrFr(\Dsphipi)$ \cite{ref:pdg2004}. The present uncertainty of about 25\%
on $\BrFr(\dsphipi)$ \cite{CLEO} thus affects many of the results
regarding \Ds mesons.

In the factorization model for two-body decay rates, it is assumed that
the transition amplitude of the process is the product of two currents 
that can be evaluated separately.
This model has been successful~\cite{rosner} in describing the measured
branching fractions and polarizations for \B\ meson decays such as
$\Bz\to\Dstarm\pip$~\cite{dstarpai}, $\Bz\to\Dstarm\rho^+$ and
$\Bz\to\Dstarm a_1^+$~\cite{dstara1}, in which the momentum transfer in 
the process is low ($q^2\simeq m_\pi^2, m_\rho^2$).
Measurements of decay rates for modes such as \bdstardsstar\
allow tests of the predictions for high $q^2$~\cite{luo}.

\section{THE \babar\ DETECTOR AND DATASET}
\label{sec:babar}
The data used in this analysis were collected with the \babar\ detector at
the \pep2\ storage ring and correspond to an integrated luminosity of
112.3\ifb\ and to $(124\pm1)\times10^6$ \BB\ pairs.
A detailed description of the detector can be found in 
Ref.~\cite{ref:babar}.

In addition to this data sample, several simulated event samples
were used in order to study efficiencies and backgrounds.
For background studies, we used a Monte Carlo simulation of \BzBzb\
and \BBpm\ events (each equivalent to an integrated luminosity of 440\ifb),
\epem\to\ccbar (230\ifb) and \epem\to\uubar, \ddbar, \ssbar (180\ifb).

\section{ANALYSIS METHOD}
\label{sec:Analysis}
The aim of this analysis is to measure the branching fractions
$\BrFr(\bdstardsstar)$ and $\BrFr(\Dsphipi)$. To accomplish this, the 
$\bdstardsstar\to(\Ds\gamma)(\Dzb\pim)$ decay is reconstructed using two
different methods.

The first method uses a partial reconstruction technique, in which only the
\Dstarm\ is fully reconstructed, and then combined with the soft photon
from the $\Dss\to\Ds\gamma$ decay, without requiring explicit \Ds
reconstruction. Denoting the measured event yield by \nDs, we can express
the \bdstardsstar\ branching fraction as:
\begin{equation}
\label{eq:brone}
\BrFr_1\equiv \BrFr(\bdstardsstar) = \Kappa\frac{\nDs}{\sum_i (\eps_i
\cdot \BrFr^{\Dz}_i)}.
\end{equation}
Here $\Kappa\equiv
[2\nBB\BrFr(\Dss\to\Ds\gamma)\BrFr(\Dstarm\to\Dzb\pim)]^{-1}$,
\nBB\ is the number of neutral \B\ meson pairs, $\BrFr^{\Dz}_i$ are the
branching fractions for the \Dz decay mode $i$, $\eps_i$ are the 
efficiencies for reconstructing and selecting the partially reconstructed
\Bo\ into a final state containing a photon, a soft pion and a \Dz\
reconstructed into mode $i$. We assume $\BrFr(\Upsilon(4S)\to\BzBzb) =
0.5$.

\bdstardsstar\ decays can also be fully reconstructed. In this paper
we focus on the \Dsphipi\ mode, where the $\phi$ meson is reconstructed
in the \Kp\Km\ channel, determining from the measured yield \nDsPhipi\ the
product of branching fractions $\BrFr(\bdstardsstar)\cdot \BrFr(\Dsphipi)$:
\begin{equation}
\label{eq:brtwo}
 \BrFr_2 \equiv \BrFr(\bdstardsstar)\BrFr(\Dsphipi) =
 \Kappa\cdot \frac{\nDsPhipi}{\BrFr(\phi\to\Kp\Km)
\sum_i (\eps'_i \cdot \BrFr^{\Dz}_i)},
\end{equation}
where $\eps'_i$ is the efficiency for detecting the fully reconstructed
\Bo, including reconstruction of $\phi\to\Kp\Km$. The branching fraction
$\BrFr(\Dsphipi)$ is measured from the ratio of the two yields:

\begin{equation}
\label{eq:brdsphipi}
\BrFr(\dsphipi)= \frac{\BrFr_2}{\BrFr_1} =
\frac{\nDsPhipi \sum_i (\eps_i \cdot \BrFr^{\Dz}_i)}
{\nDs \BrFr(\phi\to\Kp\Km)  \sum_i (\eps'_i \cdot \BrFr^{\Dz}_i)},
\end{equation}
where the factor \Kappa\ exactly drops off, and although the efficiencies
$\eps_i$ and $\eps'_i$ are in general different, many systematic
uncertainties cancel in the ratio, as will be discussed in
Sec.\;\ref{sec:Systematics}.

\section{PARTIAL RECONSTRUCTION ANALYSIS}
\label{sec:Partial}
\subsection{Signal Extraction}
We reconstruct the
$\bdstardsstar\to(\Ds\gamma)(\Dzb\pim)$ decay by combining photons in the
event with fully reconstructed \Dstarm\ mesons, without requiring
reconstruction of the \Ds from the \Dss\ decay. In order to extract the
signal, we compute the missing mass \Mmiss\ recoiling against the
$\Dstarm-\gamma$ system (all quantities defined in the center-of-mass 
(CM) frame):
\begin{equation}
\Mmiss = \sqrt{(E_\mathrm{beam} - E_{\dstar} - E_{\gamma})^2 -
({\bf p}_B - {\bf p}_{\dstar} - {\bf p}_{\gamma})^2}.
\label{eq:nostramm}
\end{equation}
The \mMiss\ distribution of signal events peaks at the nominal \Dsp\
mass\;\cite{ref:pdg2004} with a spread of about 15\MeVcc. The kinematics of
the event are not fully constrained with this procedure and one of the
decay parameters must be chosen in an arbitrary way. In particular, taking
the beam energy in the CM to be the \B\ energy, the angle between
the \B\ momentum vector and the measured \dstarm\ momentum vector can be
calculated from 4-momentum conservation in the \bdstardsstar\ decay
\begin{equation}
\cos\vartheta_{B\dstar} = -\frac{m^2_B+m^2_{\dstar}-m^2_{\dsstar}-2E_B 
E_{\dstar}}{2|{\bf p}_B||{\bf p}_{\dstar}|}.
\end{equation}
The \B\ four-momentum is therefore determined up to the azimuthal angle
around the \Dstarm\ direction. However, an arbitrary choice of this
angle (\eg\ $\cos\phi_{B\dstar} = 0$) introduces only a negligible
spread (of the order of 1.5 \MeVcc) in the missing mass distribution.

\subsection{Event Selection}
\label{sec:evSel}
Random $\Dstarm-\gamma$ combinations are suppressed requiring
$|\cos\vartheta_{B\dstar}|<1.2$.
To reject events from continuum, we require the ratio of the
second to the zeroth Fox-Wolfram moment ($R_2$) \cite{R2} to be less than 
0.3.

Candidates for \Dstarm\ are reconstructed in the $\Dzb\pim$ mode, using 
\Dzb decays to $\Kp\pim$, $\Kp\pim\pipi$ $\Kp\pim\piz$, and
$\Kos\pipi$, here listed in order of decreasing purity. The $\chi^2$
probabilities of both the \Dz\ and \Dstar\ vertex fits are required
to be greater than 1\%. The \Dstarm\ momentum in the \FourS frame   
must satisfy $1.4\GeVc < \pcms_{\dstar} < 1.9\GeVc$.
Moreover, we require the reconstructed mass of the \Dz\ to be within 3 
standard deviations $\sigma_{m_{\Dz}}$ of the nominal value $m^{\scr
PDG}_{\Dz}$\,\cite{ref:pdg2004}, and $Q_{\dstar} \equiv
m_{\dstar}-m_{\Dz}-m_{\pim}$ to satisfy $Q_{\mathrm{lo}} < Q_{\dstar} <
Q_{\mathrm{hi}}$, where the limits
$Q_{\mathrm{lo}}=4.10$ to $5.20\MeVcc$ and $Q_{\mathrm{hi}}=6.80$ to
$7.90\MeVcc$ are chosen around the nominal value $Q^{\scr
PDG}_{\dstar}=5.851~\MeVcc$ depending on the \Dz\ decay mode.
Kaon identification is required for modes $\Kp\pim\piz$ and
$\Kp\pim\pipi$. For mode $\Kos\pipi$, the invariant mass of the \pipi
from the \Kos\ decay is required to lie within 15\MeVcc\ of the nominal
\Kos\ mass and its flight length must be greater than 3\,mm.

If more than one \Dstarm\ candidate is found, for each \Dz\ decay mode
we choose the best one based on the quality of the slow pion track and on
the minimum value of
$\chi^2 =
  [(Q_{\dstar} - Q^{\scr PDG}_{\dstar})/\sigma_{Q_{\dstar}}]^2
+ [(m_{\Dz} - m^{\scr PDG}_{\Dz})/\sigma_{m_{\Dz}}]^2$, where
$\sigma_{Q_{\dstar}}$ is the measured resolution on $Q_{\dstar}$.
Finally, if candidates from different \Dz\ decay modes are present, we
select the one with the best expected purity.

Photon candidates are chosen from energy releases in the electromagnetic
calorimeter not associated with any charged tracks. In order to reduce
random associations, we reject photon candidates which form, in
combination with any other photon in the event, a \piz\ whose invariant
mass is between 115 and 155\MeVcc\ and whose momentum in the CM is greater
than 200\MeVc.
The selection of photon candidates is based on the optimization of the
statistical significance of the observed signal ($S/\sqrt{S+B}$, where $S$
and $B$ are the number of signal and background photons respectively),
using Monte Carlo events. We require a minimum photon energy in the \FourS\
CM $\Ecms$ of 142\MeV, a minimum cluster lateral moment\,\cite{LAT} of $0.016$,
and a minimum Zernike moment \cite{zernike} of order $\{2,0\}$ of $0.82$.
In about 10\,\% of the events, more than one photon is selected.
In these occurrences we choose the one that maximizes the value of a
likelihood ratio based on four photon variables ($E_{\gamma}$, \Ecms,
$N_{cry}$, LAT), where $E_{\gamma}$ is the photon  energy in the laboratory
frame and $N_{cry}$ is the number of calorimeter crystals in the
electromagnetic shower.

\subsection{Signal Yields}
\label{sec:efficiency}
The signal reconstruction efficiency is determined from a Monte Carlo
sample of \bdstardsstar\ events by performing an unbinned maximum
likelihood  fit to the missing mass distribution. The signal peak is well
described by a Gaussian probability density function
(p.d.f.), while the background, which is mainly due to random
$\Dstarm-\gamma$ combinations, is parametrized with the function
$B(\mMiss)=B_0(1-e^{-(\mMiss-\mmax)/b})({\mMiss}/{\mmax})^c$,
where \mmax\ is the end point of the missing mass distribution.
In the fit, we allow seven parameters to vary: $B_0,~b,~c$ and \mmax\
in $B(\mMiss)$, and the mean, width and area of the signal Gaussian.
We perform a single fit to all \Dz\ decay modes; the sum of the
branching fraction-weighted efficiencies for the four reconstruction modes
is computed from the number of signal events fitted in the range
$|\mMiss-m_{\Dsnosign}| < 45 \MeVcc$. The result is $\vev{\eps \BrFr} \equiv
\sum_i{(\eps_i \cdot \BrFr^{\Dz}_i)} = (5.11 \pm 0.03) \times 10^{-3}$.

We have validated the fitting technique and the method of extracting the
signal on the generic Monte Carlo sample.
The distribution of the missing mass of partially reconstructed
\Bz candidates is shown in Fig.~\ref{fig:missmasstot}a for \BzBzb 
(including signal), \BBpm, and continuum Monte Carlo events.
From the signal yield, using Eq.~\ref{eq:brtwo} we obtain
the result $\BrFr(\bdstardsstar)=(1.962 \pm 0.036)\,\%$, which is consistent
with the value (1.97\%) used in the generation of the Monte Carlo sample.

\begin{figure}[htbp]
\centering
\begin{tabular}[t]{c@{\hspace{1mm}}c}
\subfigure[ ]{
\includegraphics*[width=8cm]{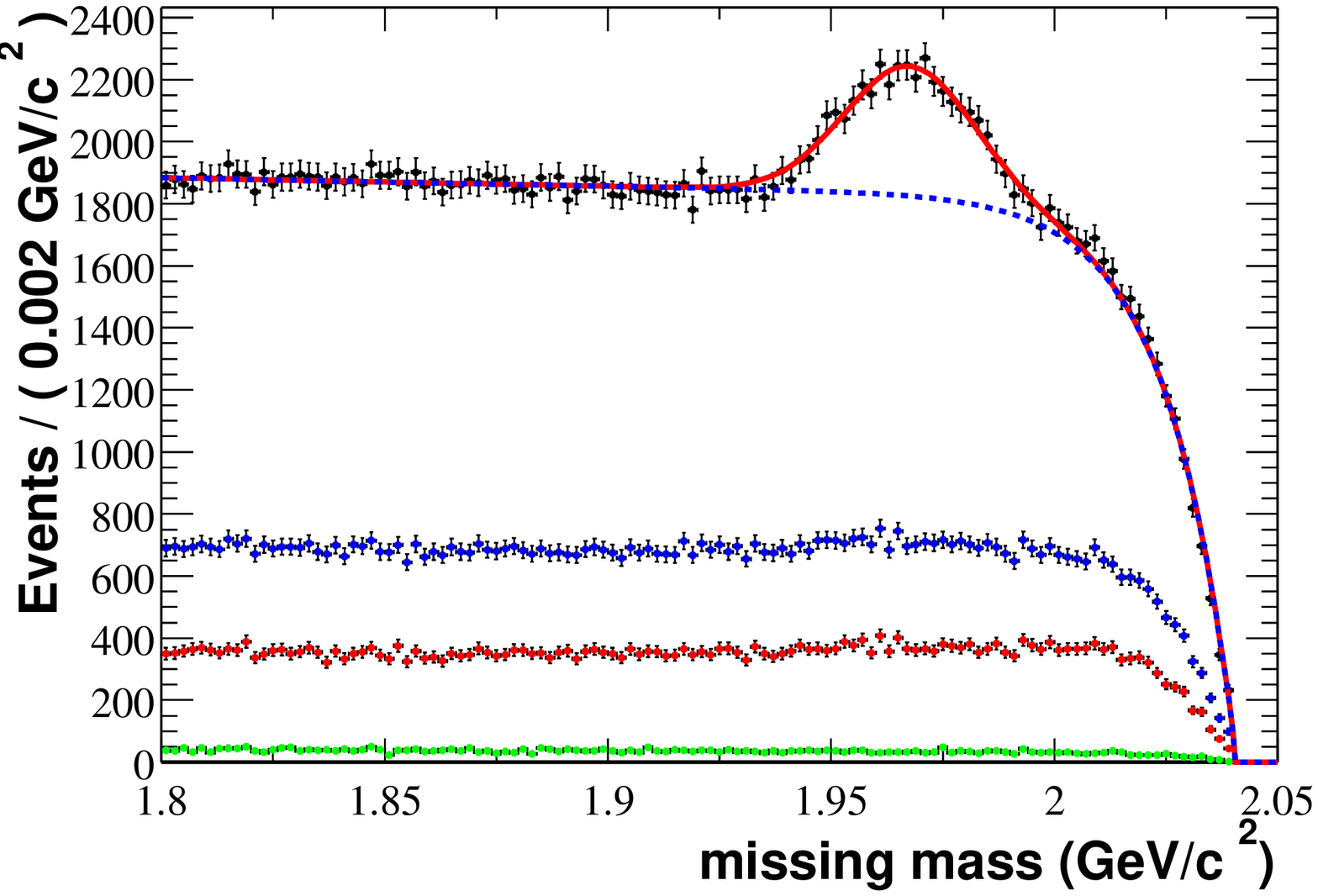} }
&
\subfigure[ ]
{
\includegraphics*[width=8cm]{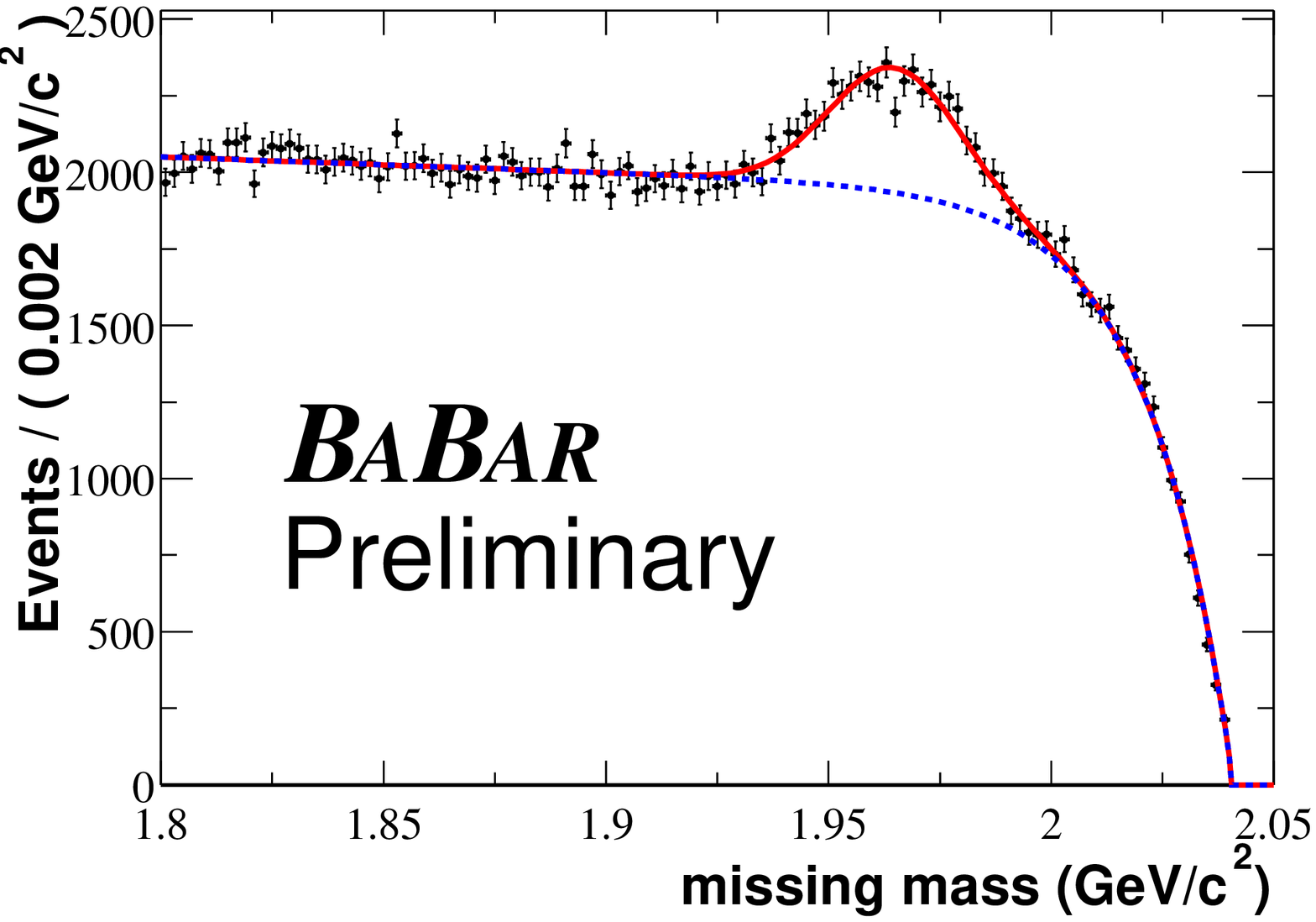}} \\
\end{tabular}
\caption{Missing mass distributions in the Monte Carlo (a)
and in the data sample (b).
The $B(\mMiss)$ background fit (dashed line) and $B(\mMiss)$+Gaussian total
fit (solid line) are superimposed. In (a) the different background
components are also overlaid; starting from the bottom: \uubar-\ddbar-\ssbar,
\ccbar, \BpBm, \BzBzb (including signal).}
\label{fig:missmasstot}
\end{figure}

Figure~\ref{fig:missmasstot}b shows the missing mass distribution in our
data sample. The same fitting procedure is used to extract the number of
signal events. In the fit we allow all parameters to vary, except the
width of the Gaussian signal, which is fixed to the value determined from
fitting the signal Monte Carlo missing mass distribution.
The result of the fit to the missing mass distribution is shown in
Fig.~\ref{fig:missmasstot}b. The signal yield is $7414 \pm 345$ events,
corresponding to a branching fraction $\BrFr(\bdstardsstar) = (1.854 \pm
0.086)\,\%$. The result is stable over different run periods.

\section{FULL RECONSTRUCTION ANALYSIS}
\label{sec:Full}
\subsection{Signal Extraction}
The full decay chain used for the $\bdstardsstar$ exclusive reconstruction 
is $\bdstardsstar\to (\Ds\gamma) (\Dzb\pim)$ with $\Dzb$ decaying
in the four modes listed above, and $\Ds\to\phi\pip\to K^-K^+\pip$.
After applying selection cuts on the $\Dss$  and $\Dstarm$ candidates, the
combination with the smallest value of $|\DeltaE| \equiv |(E_{\dstar} +
E_{\dsstar}) -E_\mathrm{beam}|$ (all quantities defined in the CM frame) is
selected.
Finally, the number of fully reconstructed \Bo\ is obtained from
a fit to the spectrum of the energy-substituted mass $\mes =
\sqrt{E^{2}_{\mathrm{beam}} - ({\bf p}_{\dstar}+{\bf p}_{\dsstar})^2}$,
where ${\bf p}_{\dstar}$ and ${\bf p}_{\dsstar}$ are the \dstarm\ and
\dsstarp\ momenta, again in the CM frame.

\subsection{Event Selection}
The selection of $\dstarm$ candidates, and most of the requirements on
photon candidates in the full reconstruction analysis,
are identical to those in the partial reconstruction.
In the full reconstruction the background level is very small. We can
therefore relax the requirement on the minimum photon energy in the
center-of-mass system $\Ecms$, thus maximizing the statistical
significance of our sample.

$\phi$ candidates are reconstructed from two charged tracks,
with at least one track satisfying stringent kaon selection criteria.
We use the helicity angle $\vartheta_K$, defined as the angle between the
kaon direction in the $\phi$ rest frame and the $\phi$ meson direction in
the \Ds\ frame, to further suppress background.

\Ds\ candidates are formed by combination with an additional track, with
charge opposite to the slow pion from the \Dstarm\ decay. A mass window of
$\pm$50\,\MeVcc\ around the nominal \Dsp\ mass\;\cite{ref:pdg2004} is
required. \Dstarm\ and \Dss\ mass constraints are finally imposed in order
to improve the \DeltaE\ resolution.

At the end of an optimization procedure based on generic Monte Carlo which
minimizes the overall statistical error and the peaking background
contribution, we require the $m_{\dsstar}-m_{\Dsnosign}$ mass difference to
be between 0.125 and 0.160 \GeVcc, the reconstructed $\phi$ mass to be
between 1.0077 and 1.0347\GeVcc, $|\cos\vartheta_{K}|> 0.35$,
$\Ecms>0.09\GeV$, and  $|\Delta E|<0.05\GeV$.

\subsection{Signal Yields}
\label{sec:EffFull}
We determine the selection efficiency fitting the \mes\ distribution of
the signal Monte Carlo sample with
a Crystal Ball function \cite{cbfunc}, defining the number of
signal events as the integral of this p.d.f. in the signal region $5.27 <
\mes < 5.29$\,\GeVcc. Summing the branching fraction-weighted efficiencies
over the four \Dz reconstruction modes (see Sec.\;\ref{sec:evSel}) yields
$\vev{\eps'\BrFr^{\Dz}} = (6.28 \pm 0.24)\times 10^{-3}$.

Monte Carlo studies show the presence of a peaking contribution due to 
real $\bdstardsstar \to (\Ds\gamma) (\Dzb\pim)$ events, where either the
\Dzb\ does not decay to the reconstructed modes, or the \Ds\ does not decay
to $\phi \pip$.

The total \mes\ distribution is fitted with the sum of a Crystal Ball
and a threshold function \cite{Argus_bkgd}, accounting for the
combinatorial background.
All parameters are allowed to vary in the fit, except the end point
of the threshold function, which is fixed to 5.29\GeVcc. 
The signal yield is obtained from the integral of the Crystal Ball
p.d.f. in the signal window, after subtraction of the peaking component
discussed above.

The fit procedure is first checked on generic Monte Carlo samples, where no
bias is found in the full reconstruction analysis technique.

Fig.~\ref{fig:fitdatafull} shows the fit to the total data sample and
the corresponding yield.
In order to subtract the peaking background from this yield, we scale
the number of peaking events previously estimated from the Monte Carlo
sample to the luminosity of the data sample. We then apply a correction 
factor to take into account the fact that the peaking
events come from real \bdstardsstar\ decays and the measured value of this
branching fraction in data (1.85\%) is slightly lower than in the Monte Carlo. 
In addition, the peaking component coming from  events with \Ds\ not
decaying to $\phi \pi^+$, is rescaled using the measured ($\Ds \to \phi
\pi^+$) branching ratio, with an iterative procedure. The resulting number
of peaking background events expected in the data sample is 35, with a
total uncertainty of 6 events which will be included in the systematic error.
After subtraction of the peaking background events, the final yield on
data is:
\begin{equation}
S_{\mathrm{tot}} = (212 \pm 19).
\label{eq:fulldatafit}
\end{equation}
From equation \ref{eq:brtwo} we therefore determine
$\BrFr_2=\BrFr(\bdstardsstar)\cdot\BrFr(\Dsphipi) = (8.71 \pm 0.78\ustat)
\times 10^{-4}$.
The result is stable over different data-taking periods.

\begin{figure}[htbp]
 \centering
\begin{tabular}{c}
  \includegraphics[width=10cm]{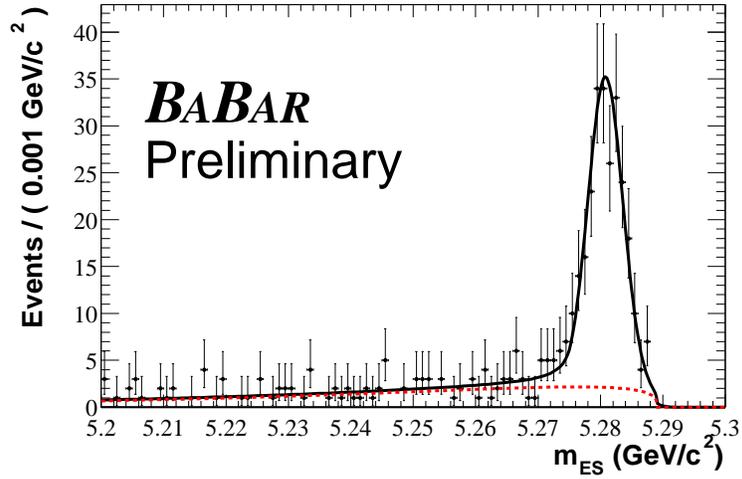}
\end{tabular}
  \caption{Total fit to the data sample. The Argus component
  parametrizing the background is also shown (dashed line).}
\label{fig:fitdatafull}
\end{figure}

\section{SYSTEMATIC STUDIES}
\label{sec:Systematics}
The main sources of systematic uncertainties on the \bdstardsstar\ 
branching
fraction measurement are listed in Table~\ref{tab:syst}.   
\begin{table}
\caption{Fractional systematic uncertainties (\%) for the \bdstardsstar\
branching fraction measurement.}
\begin{center}
\begin{tabular}{lr}
\hline
Source                       & Error $(\%)$ \\ \hline
Monte Carlo statistics       &       0.6    \\
p.d.f. modelling             &       4.5    \\
$B$ counting                 &       1.1    \\
Tracking efficiency          &       2.4    \\
Soft pion efficiency         &       1.6    \\
Vertexing                    &       2.0    \\
Photon efficiency            &       4.6    \\
Particle identification      &       0.9    \\
Polarization uncertainty     &       0.8    \\
\Dz branching fractions      &       3.2    \\
$\BrFr(\Dstarm\to\Dzb\pim)$  &       0.7    \\
$\BrFr(\Dss\to\Ds\gamma)$    &       2.7    \\
\hline
Total systematic error       &       8.6    \\ \hline
\end{tabular}
\end{center}
\label{tab:syst}
\end{table}
The Monte-Carlo-statistics uncertainty is due to the statistical error
on the efficiency determination.
The uncertainty due to the use of a fixed width for the signal Gaussian is
estimated from the spread in the fit results allowing the width to vary,
and simultaneously fixing in the fit the background shape as determined in 
several different ways. We conservatively assign an error of 4.5\,\%.
The systematic uncertainty due to tracking efficiency is 0.9\% per
track and 1.6\% for the soft pions from \Dstarm\ decays.
The systematic error due to the isolated photon reconstruction efficiency
and particle identification are evaluated using control samples.
We find a 7\% difference in the overall selection efficiency between our
samples with complete longitudinal or transverse polarization in the
\bdstardsstar\ decay. The uncertainty due to the dependence on 
polarization is computed taking into account the experimental measurement
of the fraction of longitudinal polarization,
$\Gamma_L/\Gamma=(51.9\pm5.7)\%$~\cite{BAD524}.
Finally, the uncertainties on the \Dz, \Dstarm\ and \Dss\ branching
fractions~\cite{ref:pdg2004} are propagated throughout the analysis.

\begin{table}
\caption{Sources of systematic error (\%) for the determination of
 $\mathcal{B}(\Ds\to\phi\pi^+)$.}
\label{tab:systfull}
\begin{center}
\begin{tabular}{llr}
\hline
 & Source                             & Error $(\%)$ \\\hline
Partial rec. & Monte Carlo statistics       &       0.6    \\
             & p.d.f. modelling             &       4.5    \\  \hline
Full rec.    & Monte Carlo statistics       &       3.2    \\
             & Tracking efficiency          &       2.6    \\
             & Particle identification      &       0.9    \\
             & Peaking background           &       2.8    \\
             & Combinatorial background     &       2.9    \\
             & $\BrFr(\phi\to K^-K^+)$      &       1.2    \\
\hline
             & Total                        &       7.5    \\
\hline
\end{tabular}
\end{center}
\end{table}
Some systematic uncertainties, namely $B$ counting, tracking efficiency,
soft pion efficiency, photon efficiency, particle identification,
polarization uncertainty, $\BrFr(\Dstarm\to\Dzb\pi^-)$ and 
$\BrFr(\Dss\to\Ds\gamma)$) cancel in the ratio (Eq.~\ref{eq:brdsphipi}).
All other sources are listed in Table\;\ref{tab:systfull}.
The error on peaking background is due to the Monte Carlo statistics and 
to the uncertainty on the relevant \Dzb\ and \Ds\ branching ratios; 
the error from the combinatorial background is estimated using
the \DeltaE\ sideband data sample as an alternate way of computing the
number of background events under the peak.
The error on the $\BrFr(\phi\to K^-K^+)$ is taken from \cite{ref:pdg2004}.

\section{SUMMARY}
\label{sec:Summary}
A measurement of the \bdstardsstar\ branching fraction is performed,
using data corresponding to an integrated luminosity of 112.3\ifb,
with a partial reconstruction technique. Including the systematic errors
discussed in the previous section we obtain:
\begin{equation}
  \BrFr(\bdstardsstar) = (1.85 \pm 0.09\ustat \pm  0.16\usyst)\%.
  \label{eq:ilnostro}
\end{equation}
This preliminary result is compatible with, and improves on the precision of 
previously published experimental results~\cite{ref:pdg2004,BAD524}, and
may be compared with the most recent theoretical results based on the
factorization assumption~\cite{luo}: $\BrFr(\bdstardsstar)_\mathrm{theor} =
(2.4 \pm  0.7)\%$.

The preliminary \Dsphipi\ branching fraction result is
\begin{equation}
\BrFr(\dsphipi)= (4.71 \pm 0.47\ustat \pm 0.35\usyst)\%,
\label{eq:philast}
\end{equation}
This new determination improves on published results~\cite{CLEO,dsfnal,dsbes}.

\def\NIM#1#2#3#4{{\sl NIM} #1, {\bf #2} (#3) #4.}

\section{ACKNOWLEDGMENTS}
We are grateful for the 
extraordinary contributions of our \pep2\ colleagues in
achieving the excellent luminosity and machine conditions
that have made this work possible.
The success of this project also relies critically on the 
expertise and dedication of the computing organizations that 
support \babar.
The collaborating institutions wish to thank 
SLAC for its support and the kind hospitality extended to them. 
This work is supported by the
US Department of Energy
and National Science Foundation, the
Natural Sciences and Engineering Research Council (Canada),
Institute of High Energy Physics (China), the
Commissariat \`a l'Energie Atomique and
Institut National de Physique Nucl\'eaire et de Physique des Particules
(France), the
Bundesministerium f\"ur Bildung und Forschung and
Deutsche Forschungsgemeinschaft
(Germany), the
Istituto Nazionale di Fisica Nucleare (Italy),
the Foundation for Fundamental Research on Matter (The Netherlands),
the Research Council of Norway, the
Ministry of Science and Technology of the Russian Federation, and the
Particle Physics and Astronomy Research Council (United Kingdom). 
Individuals have received support from 
CONACyT (Mexico),
the A. P. Sloan Foundation, 
the Research Corporation,
and the Alexander von Humboldt Foundation.

\end{document}